# Relativistic Perihelion Precession of Orbits of Venus and the Earth

## Abhijit Biswas and Krishnan R. S. Mani [*]

Department of Physics, Godopy Center for Scientific Research, Calcutta 700 008, India

___

**Abstract**

Among all the theories proposed to explain the 'anomalous' perihelion precession of Mercury's orbit announced in 1859 by Le Verrier, the general theory of relativity proposed by Einstein in November 1915, alone could calculate Mercury's 'anomalous' precession with a precision demanded by observational accuracy. Since Mercury's precession was a directly derived result of the full general theory, it was viewed by Einstein as the most critical test of general relativity, amongst the three tests proposed by him. With the advent of the space age, the observational accuracy level has improved further and it became possible to detect this precession for other planetary orbits of the solar system --- viz., Venus and the Earth. This conclusively proved that the phenomenon of 'anomalous' perihelion precession of planetary orbits is really a relativistic effect. Our previous papers presented the mathematical model and the computed value of the relativistic perihelion precession of Mercury's orbit using an alternate relativistic gravitational model, which is a remodeled form of Einstein's relativity theories, and which retained only experimentally proven principles and has been enriched by the benefits of almost a century-long relativity experimentation including the space age experiments. Using this model, we present in this paper the computed values of the relativistic precession of Venus and the Earth, which compare well with the predictions of general relativity and also are in agreement with the observed values within the range of uncertainty.

___

[*] E-mail: godopy@vsnl.com

## 1. Introduction

In our solar system, the potential manifestations of relativistic effect are so tiny that, in 1916, Einstein could think of only three tests in which his General relativity theory (GRT) differs from Newton's gravitational theory**:** starlight deflection, perihelion precession of Mercury, and gravitational redshift. With the advent of the space age in the 1960s, the observational accuracy level and computational capability improved further making possible the following additional tests**:** Shapiro time delay, and relativistic perihelion precession of Venus and the Earth.

## 2. Historical Background

Relativity experimentation began with renewed vigor in the sixties, and with the utilization of new space-age technologies, the accuracy of observational data, the values of planetary masses and of other astronomical constants improved considerably. As a result it became evident that the phenomena of relativistic precession exist for perihelion of the orbits of planets like Venus and the Earth, even though these are further away from the Sun and their orbits are not as highly elliptical as the orbit of Mercury. However, the centennial rate of relativistic precession for Venus and the Earth as compared to Mercury, is lower by almost an order of magnitude, and had large uncertainties in the centennial rate as derived in the seventies and the mid-eighties. However, GRT estimates for the relativistic precession for both Venus and the Earth compared well with the respective observational values within the respective high levels of uncertainty. This proved conclusively that the phenomena of 'anomalous' perihelion precession of planetary orbits is really a relativistic effect and hence cannot be explained by any of the 'ad-hoc' hypotheses [1] that were offered before Einstein presented his GRT.

## 3. Gravitational model

It may be mentioned here that the results of numerical simulations presented in this paper, are based on an alternate relativistic gravitational model titled the Remodeled Relativity Theory (RRT) as given in our earlier papers [2, 3, 4] It may be noteworthy to mention here that RRT as explained in detail in our earlier paper [4], retained only the experimentally proven principles from Einstein's relativity theories, and that RRT has been enriched by the benefits of almost a century-long relativity experimentation including the space age experiments.

RRT adopted and expressed the conservation law of momentum vector direction as its generalized law for spinning and rotational motions, from which vectorially all the expressions for inertial forces and torques (viz., centrifugal and Coriolis forces, as well as gyroscopic precessional torques for spinning tops and gyroscopes) could be derived. In fact, as mentioned below this law involving the inertial forces and torques, has been successfully used for the precision computation of planetary and lunar orbits [3, 4].

The conservation laws of energy and momentum are the most fundamental principles of the RRT. Based on almost a century-long results of relativity experiments, two fundamental principles were adopted for RRT [4]**:** one, that energy level is the underlying cause for relativistic effects and two, that mass is expressed by the relativistic energy equation as enunciated by Einstein.

Utilizing the space age ephemeris generation experience and following the methodology of nature to conserve energy and momentum, we found the reason to replace the concept of "relativity of all frames" with that of "nature's preferred frame", as explained in our earlier paper [4]. This is strongly supported by the fact that the least-squares-adjusted (LSA) astronomical constants (e.g., the planetary masses, etc.) of nature, which are an outcome of global fits done during the generation of a particular ephemeris, are consequences of not only the gravitational model, but also of the coordinate frame. In other words, the constants of nature are linked to the coordinate frame. This conclusively tells that today one has to accept the existence of only one set of the constants of nature as a concomitant of only one appropriate 'preferred frame' and the relevant orbit or orbits, linked to them. This 'preferred frame' according to the RRT has been termed as the "nature's preferred frame" [4].

Based on the few well-proven basic principles cited above, a comprehensive remodeling effort led to the RRT that has been used to consistently and successfully simulate numerically the results of all the "well-established" tests of the GRT at their current accuracy levels [2, 3], and for the precise calculation of relativistic effects observed in case of the Global Positioning System (GPS) applications, the accurate macroscopic clock experiments and other tests of the special relativity theory [4].

The mathematical model used for computation of results presented here is exactly the same as that utilized for computing the centennial rate of the relativistic perihelion precession of Mercury's orbit [3].

In this paper, we present the result of numerical simulation of the 'relativistic perihelion precession' of two more planetary orbits of the solar system --- viz., Venus and the Earth.

It may be mentioned here that no work using a similar approach could be found in the relevant literature, for the calculation of the relativistic precession of Venus and the Earth.

## 4. Numerical Simulation

Computations were done on the heliocentric equatorial-of-date co-ordinate system. JPL's (Jet Propulsion Laboratory, Caltech) ephemeris DE405 provided the data for the heliocentric positions of all planets other than the orbiting planet. Ordinary differential equations (ODE) were generated for the equations of motion based on the relevant mathematical model of RRT [3]. All the ODE's were solved simultaneously using the variable step differential equation solver, namely, the Gear's method [5]. This method is comparable to the best available in the field for astro-dynamical calculations in controlling the integration error within the specified value and has been tested against various JPL integration algorithms. The appropriate codes provided by JPL were used for extraction and interpolation of data from DE405. All the Fortran codes, have been developed by us using the above-mentioned mathematical model as presented in our earlier paper [3], other than that provided by JPL for using ephemeris data, while making use of a few of the algorithms for rotation matrices given in JPL's Technical Report [6] and in Newhall et al [7]. The results of the computation are presented and discussed in the following section.

In fact, our program can generate the ephemeris data --- that is, the epochwise position and velocity vector data of any orbiting planet of solar system while reading the corresponding position and velocity vector data of other planets and only the orbital starting epoch's position and velocity vector data of the orbiting planet, from JPL's DE405. This is done by numerically integrating the ODE's corresponding to the resultant gravitational acceleration vector. This part is similar to the JPL's ephemeris generation process in the sense that the acceleration vector is calculated from the so-called 'real forces'. But, it differs from JPL in the sense that JPL uses the general relativistic equation of motion, whereas we use the equations of motion that are based on the RRT model, and achieve comparable levels of accuracy. An additional feature of our program is the existence of a counter-checking method that numerically integrates the ODE's corresponding to the acceleration vector calculated from the so-called 'inertial forces', which are exactly equal and opposite to the corresponding 'real forces'. This enables the counter-checking of the velocity and position vectors obtained from the integration of 'real forces'. This also enables us to ensure that the conservation laws of energy, linear and angular momentum are obeyed during all stages of epochwise calculations.

## 5. Discussion of results

Using the model developed by us, the following results (Table 1) have been computed for the centennial rate of the 'relativistic perihelion precession' for the orbits of Venus and the Earth. Computations for a time frame of over two centuries were carried out. Observational results for the 'relativistic perihelion precession' have been presented from other sources for which the references have been cited.

**Table 1   Computed Values of the Relativistic Perihelion Precession in arcseconds per century compared with GRT value and Observational results**

| Orbiting Celestial body | Observational results | Uncertainty | Reference | GRT value | Relativistic perihelion precession, computed using the RRT model |
|---|---|---|---|---|---|
| Venus | 8.6247 | 0.0005 | Pitjeva (2007), [8] | 8.6247 | 8.62473 |
| Earth | 3.8387 | 0.0004 | [9, 10] | 3.8387 | 3.83868 |

For Venus, we have presented very recent observational results (that are yet to be published) derived from recent Magelan doppler data near Venus [8].

A close scrutiny of Table 1, will show that our computed values for Venus and the Earth not only agree with the respective values predicted by GRT but also agree with the recent observational results within their respective range of uncertainty.

## 6. Conclusion

It can thus be seen from the presentation of data above and discussion thereon, that RRT leads to computed values that compare well with the prediction of GRT and the recent observational results, for the relativistic perihelion precession of Venus and the Earth. Thus, the results presented in this paper can be said to confirm again the consistency of the RRT.

## 7. Acknowledgements

We thank Dr. E. V. Pitjeva, the head of Laboratory of Ephemeris Astronomy of the Institute of Applied Astronomy, Russian Academy of Sciences, for helpful clarifications, and for help in obtaining data on observational results and their uncertainties. We thank also Dr. Lorenzo Iorio for giving us encouraging help.